\documentclass[twocolumn]{aa}  

\usepackage{graphicx}

\usepackage{txfonts}

\usepackage{natbib,twoopt}
\usepackage[breaklinks=true]{hyperref} 
\bibpunct{(}{)}{;}{a}{}{,}             
\makeatletter
  \newcommandtwoopt{\citeads}[3][][]{\href{http://adsabs.harvard.edu/abs/#3}%
    {\def\hyper@linkstart##1##2{}%
     \let\hyper@linkend\@empty\citealp[#1][#2]{#3}}}
  \newcommandtwoopt{\citepads}[3][][]{\href{http://adsabs.harvard.edu/abs/#3}%
    {\def\hyper@linkstart##1##2{}%
     \let\hyper@linkend\@empty\citep[#1][#2]{#3}}}
  \newcommandtwoopt{\citetads}[3][][]{\href{http://adsabs.harvard.edu/abs/#3}%
    {\def\hyper@linkstart##1##2{}%
     \let\hyper@linkend\@empty\citet[#1][#2]{#3}}}
  \newcommandtwoopt{\citeyearads}[3][][]%
    {\href{http://adsabs.harvard.edu/abs/#3}
    {\def\hyper@linkstart##1##2{}%
     \let\hyper@linkend\@empty\citeyear[#1][#2]{#3}}}
\makeatother

\usepackage{xcolor}

\begin{document}

   \title{Asteroseismic models of the magnetic binary HD 156424}

   \author{
          C.C. Lovekin\inst{1}\fnmsep\thanks{Corresponding Author: clovekin@mta.ca}
          \and
          S. Davis
          \inst{1}
          \and
          V. Khalack
          \inst{2}
          }

   \institute{Mount Allison University, 
67 York St.
Sackville, NB, Canada
\and 
Universit\'e de Moncton
Moncton, NB, Canada
}

   \date{Received ; accepted }

  \abstract
  {HD 156424 is a hot magnetic star in the Sco OB4 association and has previously been identified as part of a binary system. Spectropolarimetric results show that the companion star is also strongly magnetic, and thus this is a rare example of a doubly magnetic hot binary. }
  {In this work, we aim to present a more detailed analysis of Transiting Exoplanet Survey Satellite data including phase variation. }
  {We find short-term phase variation consistent with an oblique magnetic rotator, as well as long-term phase variation consistent with the third element proposed previously. We performed asteroseismic modelling of the star and determined that the pulsations are most likely associated with the primary of the system. }{Our best-fit models are universally young, and we find that the star is well fitted by a model with $M$=7.5 M$_{\odot}$. }{}


   \maketitle

\section{Introduction} \label{sec:intro}

Magnetic fields have been detected in about 10\% of OB stars \citep{grunhut2017}. This detection is somewhat surprising, as OB stars lack the envelope convection responsible for generating the magnetic fields in low-mass stars.  The origin of these fields is still somewhat unclear. Existing magnetic fields may be a relic frozen in during pre-main sequence evolution, or they may be the result of binary mergers and mass transfer \citep{schneider2019,frost2024}. Nevertheless, many OB stars have been discovered to have strong ($>1$ kG) magnetic fields. This includes both HD 156424A and its companion \citep{shultz:2021}.  

The B2V star HD 156424 is known to be magnetic, with a measured mean longitudinal magnetic field of around
$-0.8$ kG \citep{alecian2014,shultz2018,shultz:2021}. In addition, the star has been observed to undergo rapid radial-velocity variation \citep{alecian2014}, and based on observations provided by the Transiting Exoplanet Survey Satellite \citep[\textit{TESS, }][]{Ricker+15}, 
\cite{shultz:2021} determined that this star is indeed both magnetic and pulsating. However, HD~156424 is a binary system, and both components are within a single \textit{TESS} pixel. As such, the pulsations could theoretically be associated with either component of the system. 

The companion star is seen in speckle observations, with an angular separation of 0.774 arcsec \citep{hartkopf1993}, decreasing to 0.3543 arcsec over a period of about 20 years \citep{tokovinin2010}. Asterometric measurements of the positions from Gaia Data Release 3 (DR3) imply a separation of 0.814 arcsec \citep{Gaia2016,Gaia2023}. Although the angular separation of the two stars is clearly changing, there are not yet sufficient data to constrain an orbit based on astrometry. The distance to this system, based on Gaia DR3 parallax for the primary star, is $1280^{+120}_{-100}$ pc, which corresponds to a projected separation of 455$^{+175}_{-70}$ AU \citep{shultz:2021}. However, it should be noted that Gaia cites different parallax measurements for the A component, $0.7810 \pm 0.0663$ mas, and for the B component,  $0.6578 \pm 0.1002$ mas \citep{Gaia2016,Gaia2023}.  These two measurements are consistent with the given uncertainties and overlap in the 0.7147-0.758 mas range.  Using this parallax gives an average distance to the system of 1360 pc, which corresponds to a projected separation of 480 AU based on the smallest angular separation of 0.3543 arcsec.

The companion (HD~156424B) has also been shown to be magnetic, with a mean longitudinal field strength of about 1.6 kG \citep{shultz:2021}.  Based on luminosity estimates from the 2.5 magnitude difference in the $y$ band \citep{tokovinin2010}, this star should be a main-sequence star of approximately 5-6 M$_{\odot}$.  \citet{shultz:2021} also found a previously undetected companion, with a mass of at least 1 M$_{\odot,}$\  making this a hierarchical triple system. This lower limit on the mass was determined by assuming that the eccentricity is zero and the radial velocity variation samples half of an orbital period, giving a period of approximately five years.

The presence of a magnetic companion makes HD 156424 one of the few known doubly magnetic hot binary stars, and the addition of pulsations make this a fascinating target for asteroseismology. Only a handful of stars have had significant magnetoasteroseismic analysis performed \citep[e.g. HD~43317;][]{buysschaert2017,buysschaert2018,lecoanet2022}. The few magnetoasteroseismic studies of magnetic pulsating hot stars thus far is largely due to too few suitable stars being detected.  Although \citet{neiner2021} identified a number of candidate hot magnetic stars, including an estimated 25 that show pulsation signals in \textit{TESS} data, no analyses of pulsating magnetic stars have been performed as yet. These pulsating magnetic stars, especially in binary systems, thus present us with a rare opportunity to investigate the structure and properties of magnetic stars, and potentially improve our understanding of the origin of these fields and their effect on stellar evolution.

Magnetoasteroseismology has been used successfully in other regions of the HR diagram.  For example, \citet{li2022} discovered magnetic fields in the deep interiors of three red giant stars, and ten more were detected by \citet{deheuvels2023}. The coupling between gravity and pressure modes forms mixed modes, allowing stars to be probed both near the surface and in the core as rotation and magnetism cause splitting of the frequencies \citep{mathis2023,rui2023,loi2020,loi2021,mathis2021,lecoanet2022}. This is usually done for red giants; however, main-sequence (MS) pulsators can have g modes that extend to the surface, allowing them to be examined similarly \citep{rui2023}. In this way, \citet{lecoanet2022} used asteroseismology to determine the interior magnetic-field strength of the slowly pulsating B star HD~43317. 

Other types of variability have been detected and studied in hot stars.  Recently, \citet{shen2023} performed a variability study of a large sample of hot magnetic stars, identifying several rotating variable stars. Their sample of 118 stars included ten with coherent pulsation, including HD 156424, and all ten stars were previously known to be pulsating. The focus of their work was on the stochastic low-frequency (SLF) variability of the stars in their sample. They were able to fit several parameters for the stars in their sample and concluded that SLF variations are common in hot magnetic stars. The origin of this variation is not clear, and it may arise from internal gravity waves \citep[e.g.][]{bowman2019,bowman2020} or from sub-surface convection zones \citep[e.g.][]{cantiello2021}. In either case, the nature of the SLF variability is significantly different from the $p$ modes typical of $\beta$ Cephei stars such as HD~156424. When \citet{shen2023} fitted the SLF variability in HD~156424, they found a characteristic frequency of $\nu_{char} = 0.21 \pm 0.07$ d$^{-1}$, which is much lower than the frequencies studied in this work. As such, we do not expect the presence of SLF variability in HD~156424 to have a significant effect on the magnetoasteroseismology of $p$ modes.

In this work, we further investigated the pulsation properties of HD 156424, including full asteroseismic modelling of both the primary and the secondary component.   The TESS observations and analysis are provided in Section \ref{sec:Obs}, and our models are discussed in Section \ref{sec:methods}. The results of our modelling are outlined in Section \ref{sec:results}, and we summarise our results in Section \ref{sec:conclusions}.

\section{Photometric analysis}\label{sec:Obs}

\begin{figure}
    \includegraphics[width=\columnwidth]{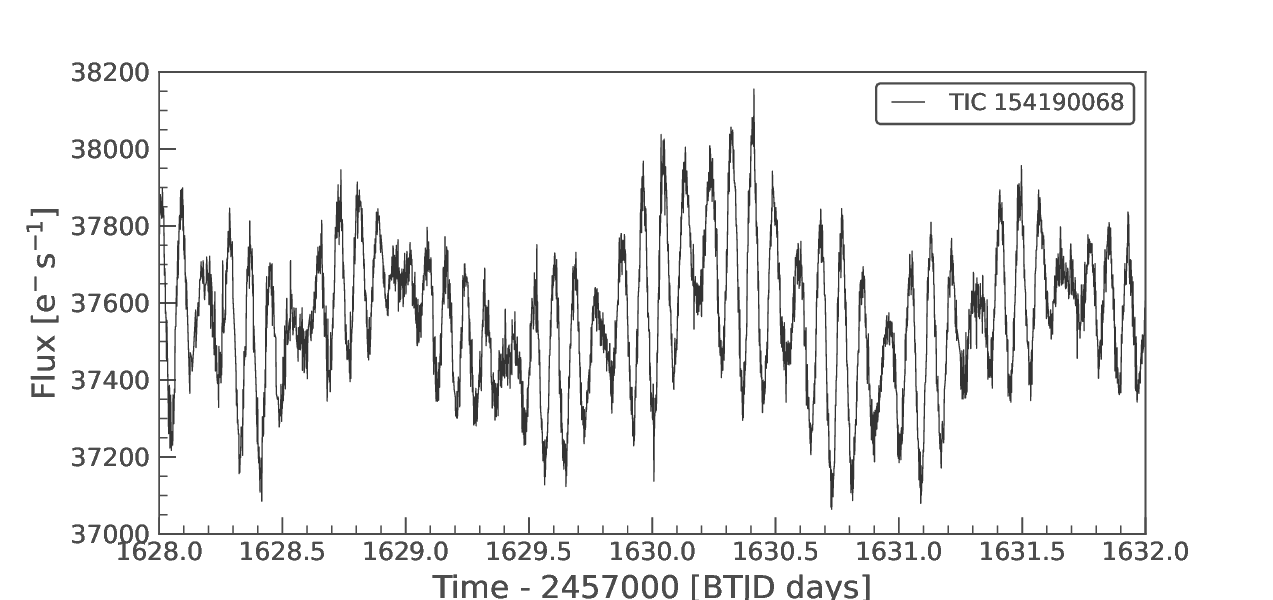}
    \caption{\label{fig:lightcurve}Four-day segment of the \textit{TESS} light curve derived from sector 12 for HD~156424.}  
\end{figure}

The star HD~156424 has been observed with \textit{TESS} in sectors 12, 39, 66, 91, and 93.  The light curve is clearly variable, as shown in Figure \ref{fig:lightcurve}.  In the original analysis by \citet{shultz:2021}, they used TESS data from sector 12 to find 11 significant frequencies, six of which were considered possible harmonics or combination frequencies.  The strongest of these, at 11.2 c d$^{-1},$ was also detected in spectroscopic data. For clarity, we use $\tilde{f}$ when referring to frequencies found in \citet{shultz:2021} and $f$ to refer to frequencies found in this work.

We extracted photometric light curves from \textit{TESS} images using the Python code developed by Jonathan Labadie-Bartz (for details, see \citealt{Labadie-Bartz+22}). This method uses the \textit{TESScut} software \citep{Brasseur+2019} and the Mikulski Archive for Space Telescopes (MAST)\footnote{https://mast.stsci.edu/tesscut/}  to cut out 24$\times$24 pixel images of HD~156424 and surroundings from the full-frame images, along with the Lightkurve Python package \citep{lightkurve} to measure the corresponding flux in the \textit{TESS} pass band. 
We implemented the principal component analysis (PCA) detrending method with five regressors to remove the sky background from the 
light curves for all five 
studied sectors. 
The light curves from sectors 66, 91, and 93 with $t_s$=158~s cadence were cleaned for outliers showing more than four-sigma deviations from the mean value. The mean and standard deviation were calculated for a sliding time window of 15800~s in width. 
The same procedure was used to clean outliers in the light curves from sectors 12 (1426~s cadence) and 39 (475~s cadence), employing sliding time windows of 42780~s and 14250~s in width, respectively, at the chosen threshold of three sigma.
To study periodic variability of the derived light curves we employed the Lomb--Scargle periodogram \citep{VanderPlas18}, which allowed us to detect presumably 
significant periodic signals, and the code Period04 \citep{lenz04} to measure frequencies, amplitudes, and phases for these signals. Period04 offers a set of well-tested powerful tools for uncertainty evaluation, and the
corresponding error bars were calculated with this code 
using Monte Carlo simulations with 1000 processes. The results for each sector are summarised in Table \ref{tab_freq12}.

We also analysed the combined light curve, with results shown in Table \ref{tab_freq_all}. When analysing the full light curve, we found a number of closely spaced frequencies that are not present in the individual sector data.  For example, frequencies were detected at 10.8617, 10.8635; 13.7751 d$^{-1}$, 
and 22.5639, 24.64 d$^{-1}$,
all with significant amplitudes. Several of these frequencies are very close to the detected frequencies in Table \ref{tab_freq_all}, and they likely arise from variability, as discussed below. 
The error bars shown in Table \ref{tab_freq_all} were derived using Monte Carlo simulations with 5000 processes, and the signal-to-noise ratio (S/N) was calculated considering two points in the frequency domain on each side of the studied frequency and using the discrete Fourier transformation (DFT) of the original data 
\citep[for details, see][]{lenz04}.

\begin{table*}
\begin{center}
\caption{Frequencies extracted from sectors 12, 39, 66, 91, and 93.  
\label{tab_freq12} }
\begin{tabular}{llllll}
\hline\hline
ID & ID  & Frequency &Sector & Amplitude & S/N  \\
 & (Shultz et al) &  (d$^{-1}$) & & (mmag) &   \\
   \hline
 $f_1$ & $\tilde{f_2}$ &  0.7171 $\pm$ 0.0003 & 12 & 2.82 $\pm$ 0.04 & 11.3 \\
 & & & 39 & 2.96 $\pm$ 0.02 & 11.5 \\
 & & &  66 & 2.923 $\pm$ 0.009 & 14.9 \\
 & & & 91 &  2.76 $\pm$ 0.02 &  5.3 \\
 & & & 93 & 2.66 $\pm$ 0.02 &  4.6 \\
 \hline
$f_2$ & $\tilde{f_1}$ &  11.2069 $\pm$ 0.0003 & 12 & 2.66 $\pm$ 0.04 & 12.3 \\
& & & 39 & 2.91 $\pm$ 0.02  & 18.0 \\
& & & 66 & 2.930 $\pm$ 0.009 & 16.8 \\
& & & 91 & 2.92 $\pm$ 0.02 &  4.7 \\
& & & 93 & 2.96 $\pm$ 0.02 &  4.3 \\
\hline
$f_3$ & $\tilde{f_3}$ &13.7756 $\pm$ 0.0005 &12 &  1.64 $\pm$ 0.04 &  15.5 \\
& & &  39 & 2.16 $\pm$ 0.02 &  17.9 \\
& & &  66 & 1.576 $\pm$ 0.009 & 18.2 \\
& & & 91 & 1.42 $\pm$ 0.02 & 5.7 \\
& & & 93 & 1.62 $\pm$ 0.02 &  5.2\\
\hline
$f_{4,5}$ & $\tilde{f_4}$ & 10.8626 $\pm$ 0.0013 & 12 & 1.44 $\pm$ 0.04 &  6.7  \\
& & & 39 & 0.16 $\pm$ 0.02 & 1.0 \\
& & & 66 &  0.775 $\pm$ 0.009 & 4.5 \\
& & & 91 & 1.59 $\pm$ 0.03 & 2.5 \\
& & & 93 & 1.52 $\pm$ 0.02 & 2.2\\ 
\hline
$f_6$ & $\tilde{f_7}$  &  22.36 $\pm$ 0.01 &12 &  0.11 $\pm$ 0.04 & 2.7 \\
& & & 39 & 0.10 $\pm$ 0.02  & 5.3 \\
& & & 66 &  0.120 $\pm$ 0.009  & 8.2 \\
& & & 91 & 0.18 $\pm$ 0.02 & 3.5 \\
& & & 93 & 0.11 $\pm$ 0.02 & 3.2\\
\hline
\hline

\end{tabular}
\end{center}
\end{table*}
  
\begin{figure}
    \includegraphics[width=\columnwidth]{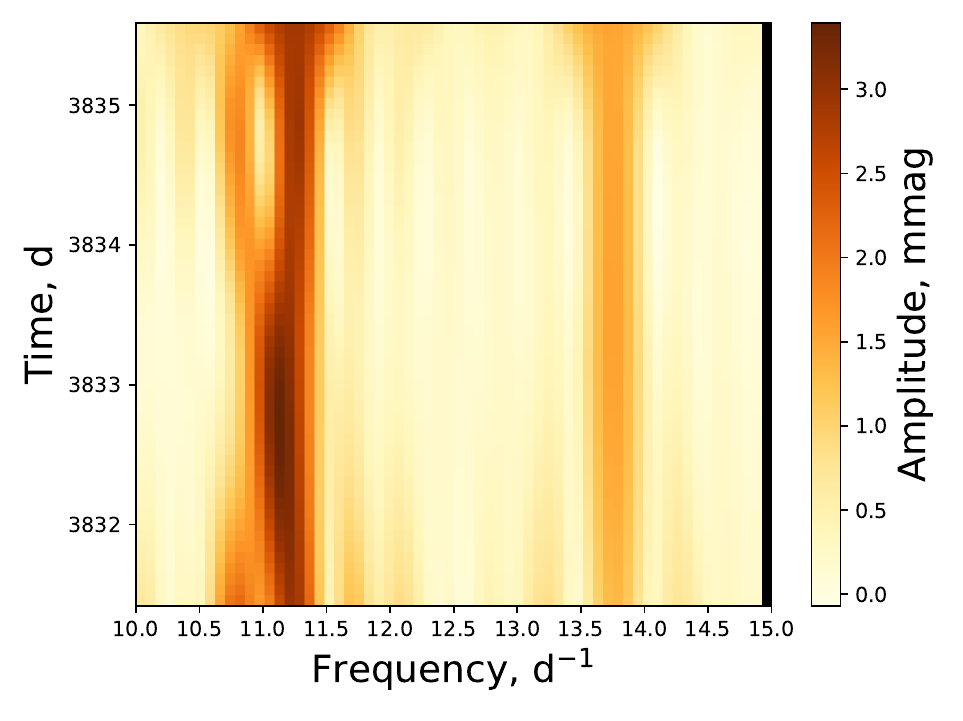}
    \caption{\label{fig:river}Dynamic-frequency spectrum plot showing the amplitude of the pulsation spectrum in the 10--15 d$^{-1}$ 
    range over \textit{TESS} sector 93. Each horizontal slice corresponds to a 2.9-day segment
    of the light curve. Each segment here (except for the first and last) overlaps with half of the previous segment and the half of the next segment.
    The variability in the amplitude of $f_{4,5}$=10.86 d$^{-1}$ is particularly prominent in this sector. }
\end{figure}

Our results are broadly consistent with the earlier analysis by \citet{shultz:2021}.  The three strongest frequencies in Tables \ref{tab_freq12} and \ref{tab_freq_all} matched the $\tilde{f}_1$, $\tilde{f}_2$, and $\tilde{f}_3$ frequencies identified by \citet{shultz:2021} at 11.2, 0.71, and 13.7 d$^{-1}$, respectively.  The next strongest frequency, $f_{4,5}$ at 10.8 d$^{-1}$ \citep[$\tilde{f}_4$ in][]{shultz:2021}, is quite variable, with an amplitude dropping from 1.44 mmag in sector 12 to 0.16 mmag in sector 39, increasing to 0.77 mmag in sector 66 and to 1.59 mmag in sector 91; it remains almost at the same level of 1.52 mmag in sector 93. We also found some variability in the amplitude of $f_3$.  As can be seen in Figure \ref{fig:river}, $f_{4,5}$ is quite variable, even within an individual sector of TESS data. This variability in amplitude and frequency gives rise to the close pairs of frequencies observed in the combined TESS light curve, such as the frequency at 13.7751 d$^{-1}$ and $f_3$, or $f_{4,5}$ and the frequencies at 10.8617 and 10.8635 d$^{-1}$.  

We did not detect several of the proposed combination frequencies previously identified by \citet{shultz:2021}, including $\tilde{f}_5$, $\tilde{f}_6$, or $\tilde{f}_{10}$. 
We also do not see a significant frequency at 16.187 c d$^{-1}$ (their $\tilde{f}_9$). 
Since $\tilde{f}_5$ is identified as the second overtone of $\tilde{f}_2$, (our $f_{1}$), it seems unlikely that our detrending methods could remove this frequency. The frequencies $\tilde{f}_6$, $\tilde{f}_9$, and $\tilde{f}_{10}$ are all greater than 11 d$^{-1}$, and it is unlikely these would be removed by de-trending unless they originated in nearby stars. The frequencies missing from our dataset correspond to the frequencies from \citet{shultz:2021} with the lowest S/N=4-5.
In addition, while we do detect their frequency $\tilde{f}_7$ at 22.36 d$^{-1}$, we find that it is not significant in sectors 12, 91, or 93. This frequency is present at higher significance in sectors 39 and 66, as the noise level in this part of the spectrum is lower.
It is found in the combined light curve, although its amplitude 
is considerably lower than the amplitudes of other significant frequencies (see Table~\ref{tab_freq_all}). A nearby frequency at 24.64 d$^{-1}$ in the combined light curve, corresponding to $\tilde{f}_8$ in \citet{shultz:2021}, also has an amplitude and significance much lower than those of the other significant frequencies.

\begin{table*}[h] \begin{center} \caption{As for Table \ref{tab_freq12} for frequencies extracted from the combined light curve. }
\label{tab_freq_all} \begin{tabular}{lllll} \hline
 ID & Frequency &Amplitude &Phase & S/N \\
 &  (d$^{-1}$) & (mmag) & &  \\  \hline
 $f_1$ &  0.7167423 $\pm 8\times10^{-7}$ & 2.862 $\pm$ 0.005 & 0.3238 $\pm$ 0.0005 & 19.2 \\ 
 $f_2$ & 11.2067557 $\pm 4\times10^{-7}$ & 2.909 $\pm$ 0.007 & 0.1084 $\pm$ 0.0004 & 19.9 \\ 
 $f_3$ & 13.775093 $\pm 9\times10^{-6}$  & 1.74  $\pm$ 0.03  & 0.811 $\pm$ 0.005 & 25.3 \\ 
 $f_4$ & 10.8617 $\pm$ 0.004  & 0.78 $\pm$ 0.07 & 0.74 $\pm$  0.16  &  5.4 \\ 
 $f_5$ & 10.8636 $\pm$ 0.001  & 0.59 $\pm$ 0.03 & 0.72 $\pm$  0.14  &  4.1 \\ 
 $f_6$& 22.36594 $\pm$ 0.00002 & 0.113 $\pm$ 0.009 & 0.29 $\pm$ 0.01 & 9.9 \\  
 $f_7$& 24.64 $\pm$ 0.02 & 0.04 $\pm$ 0.01 & 0.71 $\pm$ 0.22 & 4.2 \\  
\hline \end{tabular} 
\tablefoot{Frequencies $f_4$ and $f_5$ are the same within errors and are considered to correspond to $f_{4,5}$ in Table \ref{tab_freq_all}.}
\end{center} \end{table*}

\begin{figure}
    \includegraphics[width=\linewidth]{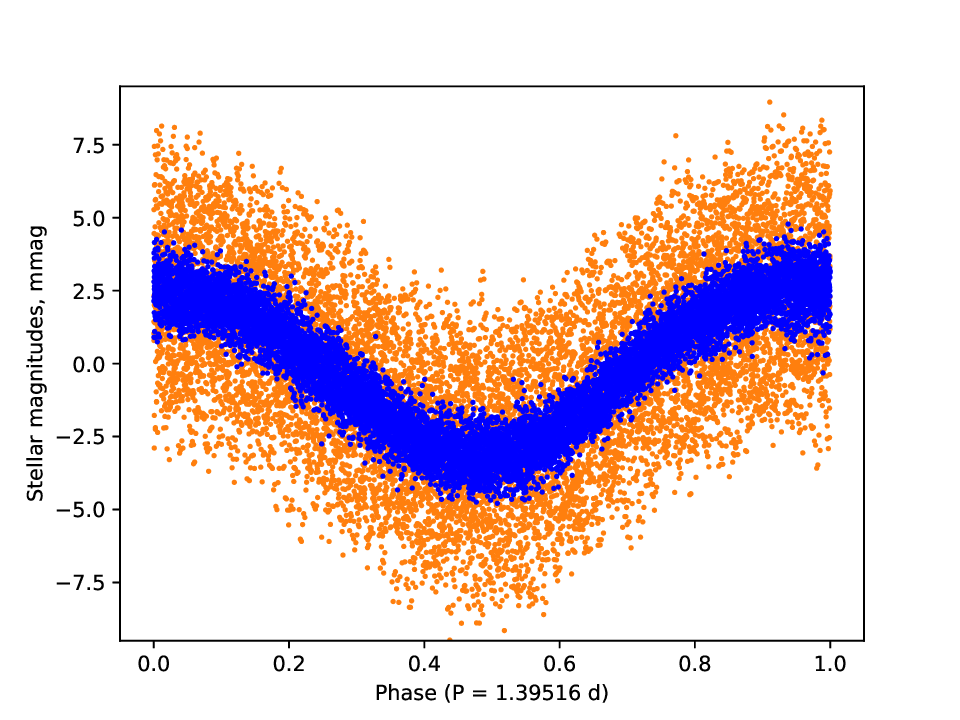}
    \caption{\label{fig:phased}Sector 66 light curve of HD 156424, phased on a period of 1.3954 d (orange points) and corresponding to the frequency $f_1$ = 0.7167 d$^{-1}$. The orange points show the original light curve, while the blue points show the phased light curve after pre-whitening all frequencies $>$ 10 d$^{-1}$.}
\end{figure}

\citet{shultz:2021} proposed that the frequency at 0.716 d$^{-1}$ is either the rotation frequency, or, based on previous measurements \citep{shultz2018}, that this could be 2$f_{rot}$.  Our results show that the light curve can be phased on the frequency $f_1$ = 0.7167 d$^{-1}$, which is consistent with rotational modulation, as shown in Figure \ref{fig:phased} for sector 66.  As a further test, we pre-whitened the light curve of all frequencies greater than 10 d$^{-1}$ and then phased the resulting light curve on $f_1$. The resulting light curve clearly shows 
only a single wave, suggesting that $f_1$ does indeed correspond to the rotation frequency.

If \citet{shultz2018} is correct and the rotation period of this star is about 2.8 d, the rotational modulation needs to be a nearly perfect double wave, as there is no evidence of a significant frequency at $f_1/2$ in the periodogram derived for any of the studied sectors or for the combined light curve.  
Indeed, in their analysis, \citet{shultz2018} noted that the variation in the magnetic field is small compared to the error bars, and the FAP of the $\langle B_z\rangle$ peaks in the period spectrum is quite low.  
However, as discussed below (see Section \ref{sec:phases}), a frequency of $f_1/2 \approx 0.35$ d$^{-1}$ arises in other places, which may support the hypothesis of a more slowly rotating star. 

The rotation frequency proposed by \citet{shultz2018} of $0.35~\rm{d}^{-1}$ 
does correspond to the separations between $f_2$ and $f_{4,5}$ 
(see Table~\ref{tab_freq12}). Of these two frequencies, the amplitude of $f_2$ is very consistent over the five observed sectors. At the same time, the very close frequencies $f_4$ and $f_5$ only appear in the combined light curve, and each individual sector only shows a single significant frequency near 10.86 d$^{-1}$. The amplitude of this frequency is also extremely variable, almost disappearing entirely in sectors 29 and 66. As discussed in Sect. \ref{sec:shortterm}, there does appear to be beating between $f_2$ and $f_{4,5,}$ with a beat frequency that is approximately $f_1/2$. While $f_2$ shows the smooth phase variation expected from beating, the phase variation in $f_{4,5}$ is more complex and may have other origins. 
It seems that the relationship between frequencies $f_{4,5}$ and $f_2$ is complex, and the correspondence between the rotation frequency and the separation is a coincidence rather than rotational splitting.

Based on our analysis of the TESS data for HD~156424, we restricted our frequency modelling in Section \ref{sec:results} to the four significant p-mode frequencies of $f_2, f_3, f_{4,5}$, and $f_6$ found in our data, as given in Table \ref{tab_freq_all}. We excluded 
$f_7$ from the analysis since it does not appear in any of the individual sectors (see Table~\ref{tab_freq12}), and we assumed that $f_1$ is the rotation frequency of the star.

\section{Phase modulation}
\label{sec:phases}

As discussed above, HD 156424 is a known binary star, with the two components separated by about 0.35'' --0.77'' \citep{hartkopf1993,tokovinin2010}.  \citet{shultz:2021} also proposed the existence of a third element in the system based on perturbations in the radial-velocity variability.  This companion is thought to have 
a much lower mass, with an upper limit of about 1 M$_{\odot}$ assuming an orbital period of about five years.  Our observed TESS data span a total period of about four years, and so we checked our data for evidence of this low-mass companion.  Following the method of \citet{murphy2014}, we divided the light curve into bins and calculated the Lomb--Scargle periodogram for each bin using a fixed reference point for the phase each time. We started with bins of 0.64 d to study short-term variations within each sector. Then, based on the resulting variation, we re-binned the data to 5.8 d intervals to address the longer term variations in phase. 

\subsection{Short-term variability}
\label{sec:shortterm} 

Within each sector, we divided our data into 0.64-day bins and calculated the phase and amplitude of each frequency and looked at the variation in phase with time. The bin size was experimentally chosen such that it would be smaller than $1/f_1$= 1.3954~d, cover several periods of the detected higher frequencies, and provide relatively small error bars for the phase measurements. 
We found significant short-term variability in all frequencies.  This is shown for $f_2 = 11.207$ d$^{-1}$ and $f_{4,5} = 10.863$ d$^{-1}$ in Figure \ref{fig:phase_f5}. In most cases, the phase variability clearly appears to be periodic, with periods that were consistent across all five sectors of the TESS data. Both $f_2$ and $f_{4,5}$ show a consistent variation of phases with a period of approximately 2.889 d (0.346 d$^{-1}$), which is close to 2.79 d (and corresponds to frequency $f_1/2$); however, there is substantially more scatter in the phase plot if $f_1$ 
is used to phase the data. 

It is possible that the difference between the period of the phase variability and the rotation period is due to differential rotation between the driving regions for the $f_2$ and $f_{4,5}$ modes and the surface of the star. In our models, the driving region for these modes is quite close to the surface; it is concentrated in the outer 2.5\% of the stellar radius, with a peak in differential work around 0.975 R$_*$. If the 2.889 d period corresponded to rotation at this depth while the observed 0.1767 d$^{-1}$ rotation frequency = 1.39516 d rotation period corresponds to surface features, the sub-surface region of the star would be rotating more slowly than the surface. This is true even if the true rotation period is $f_1/2$, in which case the rotation period would be 2.790 d. It is hard to imagine effective angular transport mechanisms in the presence of a magnetic field that could produce this kind of profile.

Differential surface rotation has been previously detected in magnetic Am stars \citep[e.g.][]{blazere2020} as a result of magnetic shear across the surface, and it is possible a similar effect is seen here. To our knowledge, this would be the first example of differential rotation observed in a magnetic B star.  In this case, the differential rotation would be between the pole and the equator. Rotation is known to confine pulsation modes to the equatorial regions of the star \citep[e.g.][]{reese2022}, with the effect increasing with the speed of rotation. Even in slowly rotating stars such as HD~156424, there is expected to be some level of confinement. This means the 2.889 d period would most likely be the equatorial rotation period. The observed rotational modulation at $f_1$ would then be the result of surface features closer to the poles. {However, this is entirely speculative. There are some arguments that strong fossil magnetic fields are not expected to be compatible with differential rotation \citep[see e.g.][and references therein]{keszthelyi2023}. For the rotational modulation to be the result of features near the pole, HD~156424 must be observed at high inclination, while our results below suggest the star is observed at relatively low inclination. Combined, this makes it unlikely that differential rotation can explain the difference in observed period.

The phase variation in $f_2$ is smooth when phased over a period of 2.889 d. This period was estimated using a DFT analysis of phases found for $f_2$ using the 0.64-day bins. This smooth variation is expected to arise from beating between frequencies. 
The beat frequency, 0.346 d$^{-1}$, is approximately the same as the difference between $f_2$ and $f_{4,5}$. However, while $f_2$ shows smooth variation with phase, $f_{4,5}$ does not (see Figure~\ref{fig:phase_f5}).  

Over each cycle, the phase of $f_{4,5}$ rises from zero to a maximum and then drops back to zero before repeating, as shown in Figure \ref{fig:phase_f5}. This variation is consistent across all five sectors and clearly phases with a period of 2.889~d. This is consistent with the type of variation seen in an oblique pulsator model \citep{bigot2002, Bigot2011}. If the strength of the magnetic field is comparable to the perturbation induced by rotation, the pulsation axis of the dipole modes will not align with either the magnetic or the rotation axis.  As a result, the observed amplitude of the mode varies as the star rotates.  The displacement vector of the dipole mode traces out an ellipse over the course of a pulsation cycle. The plane of the ellipse is defined by the vectors of the magnetic field and the rotation. The reader should refer to \citet{Bigot2011} for more details concerning the geometry of the model.  

\citet{Bigot2011} showed that the amplitude and phase of the star pulsating at a frequency, $\omega,$ are expected to vary as
\begin{equation}
\frac{\delta L}{L}(t) \propto R(t)\cos(\Omega t - \Psi(t))
,\end{equation}
where
\begin{equation}
    R(t) = \sqrt{A^2 + B^2}
\end{equation}
and 
\begin{equation}
\Psi(t) = \arctan\left(\frac{B}{A}\right).
\end{equation}
The functions $A$ and $B$ are defined as
\begin{eqnarray}
    A &=& \cos\psi\left(\cos\gamma\cos i +\sin\gamma\sin i \cos(\Omega t)\right),\\
    B &=& \sin\psi\sin i \sin(\Omega t)
,\end{eqnarray}
where $i$ is the inclination of the system, $\Omega$ is the angular rotation velocity of the star, $\gamma$ is the angle between the rotation axis and the plane of the ellipse traced out by the displacement vector of the dipole mode, and $\psi$ is the polarisation axis defined by the ratio of the two axes of the ellipse.

\begin{figure}

    \includegraphics[width= 3.5 in]{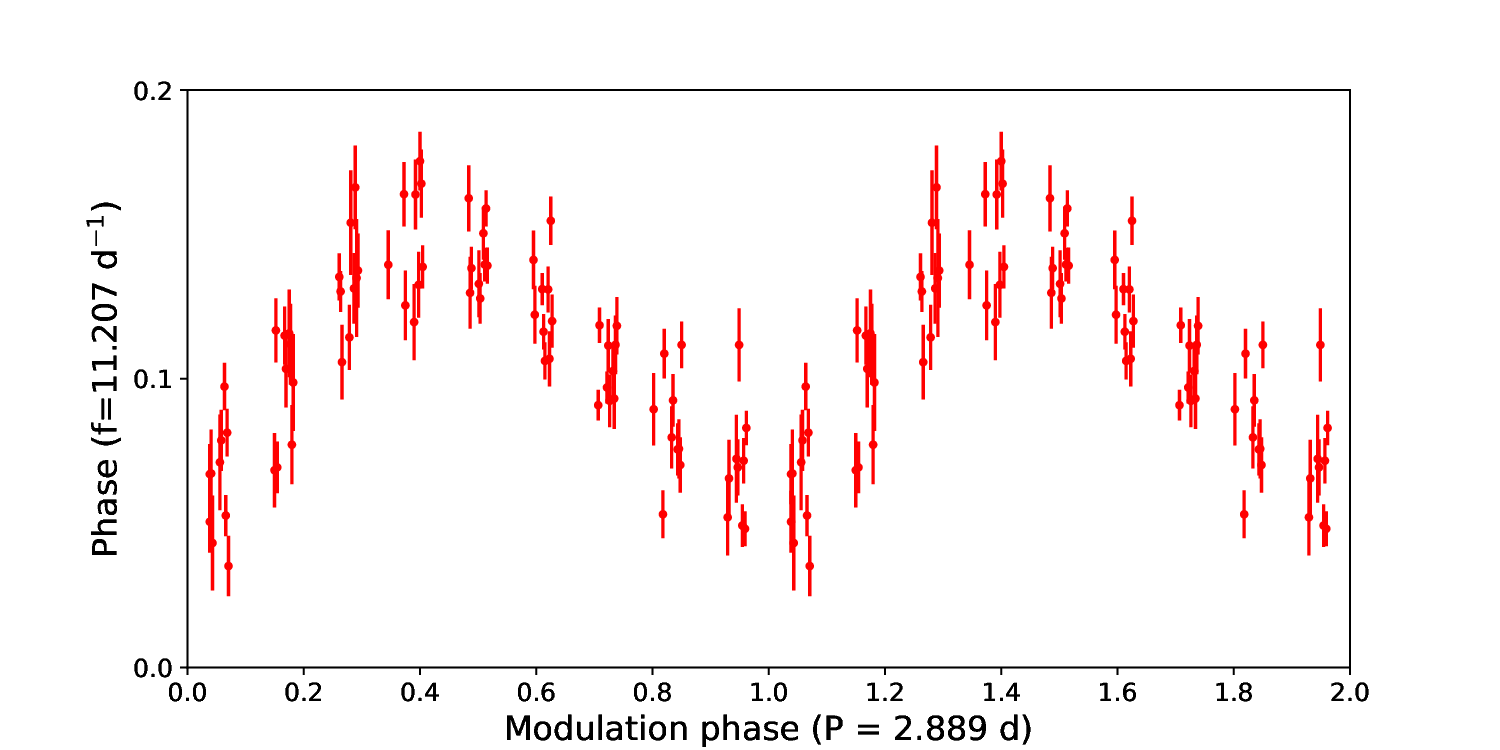} \\
    \includegraphics[width= 3.5 in]{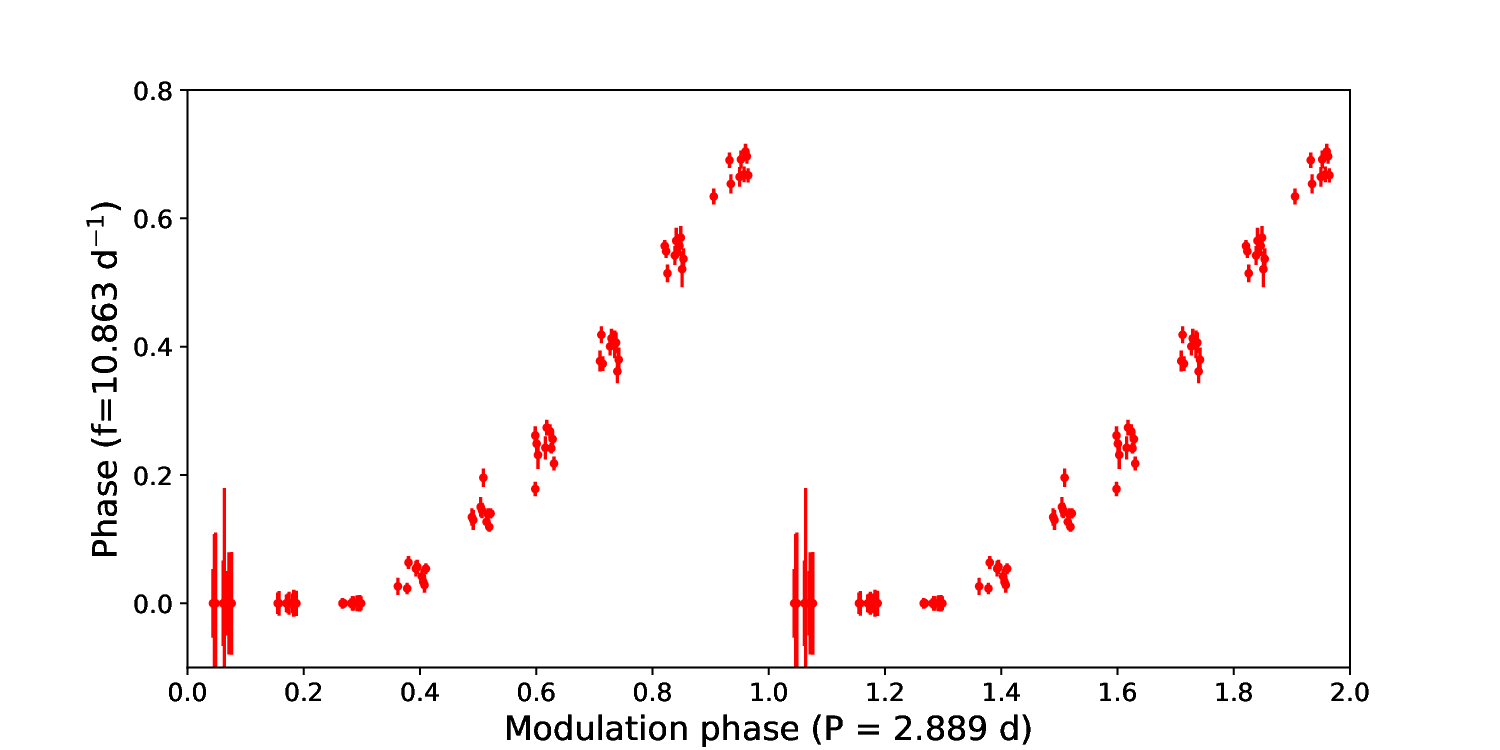} 

    \caption{\label{fig:phase_f5}
    Top panel: Phase variation of $f_2$ during sector 66 (see Table~\ref{tab_freq12}), phased over two periods of 2.889 d. 
    Bottom panel: Same as top but for $f_{4,5}$. 
    Phases were calculated in 0.64-day bins for sector 66. These variations are periodic with a frequency of 0.346 d$^{-1}$, which is approximately $f_1/2$; this is in agreement with the rotation period of the star from magnetic data \citep{shultz2018}. The rise in phase and the sharp drop over each cycle can be clearly seen in the right panel. The zero phase was chosen separately for each frequency to ensure that the phase minimum occurs at the zero point of the modulation phase.}
\end{figure}

We attempted to fit our data for the phase variation of $f_{4,5}$ to the oblique pulsator model derived in \citet{Bigot2011}, which assumes the observed pulsation is a dipole mode. We used the Python package \texttt{dynesty} \citep{dynesty,speagle2020,skilling2004} to explore the parameter space with nested sampling. We ran models allowing the three main variables in \citet{Bigot2011} to be free parameters: $\psi$, which measures the ratio of the axes of the elliptical motion; $\gamma$, which is the inclination of the $X$-axis of the ellipse with respect to the rotation axis; and $i$, the angle between the rotation axis and the line of sight.

 We found that to reliably fit the data, we needed to set $\Omega$ = 1.0874 rad d$^{-1}$, which corresponds to a period that is approximately twice the phase period shown in Figure \ref{fig:phase_f5}, or four  
times slower than the proposed rotation period of the star ($f_1$). It is not clear how to account for this discrepancy in rotation period. The other fit parameters were best fit at values of $\psi = 0.14^{+0.01}_{-0.08}$ rad, $\gamma = 1.21^{+0.31}_{-0.77}$ rad, and $i = 1.08 ^{+0.40}_{-0.82}$ rad, which correspond to $\psi = 7.5^{\circ}$, $\gamma = 65.5^{\circ}$ and $i = 57.8^{\circ}$. We repeated this process for sector 66, which also shows the same behaviour. Our fits in that sector were similar and agree with the sector 12 values within errors. The comparison between our phase data and the resulting fit is shown in Figure \ref{fig:oblique}. 

We found that there were clear non-linear correlations between the variables, as shown in Figure \ref{fig:corner}. In particular, the best-fit values of $\gamma$ and $\psi$ are both strongly dependent on the inclination of the system. This is not surprising, and we expect to see some degeneracy. Changing the orientation of the rotation and magnetic axes will change both $\gamma$ and $\psi$ relative to the observer, and it should be possible to change the inclination to produce a similar effect.

\begin{figure}
\includegraphics[width=\linewidth]{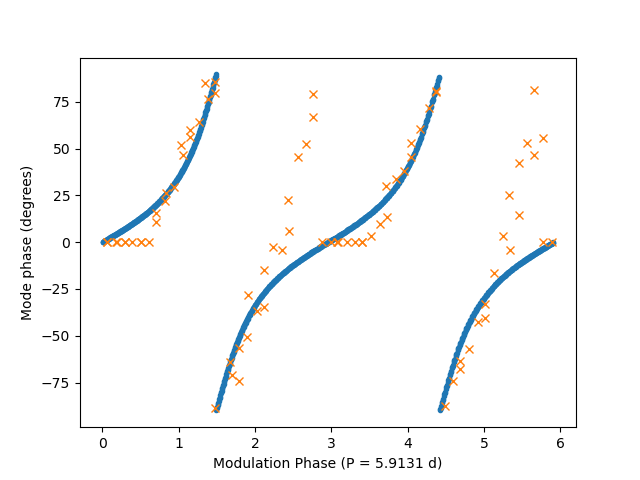}
    \caption{\label{fig:oblique}Comparison between the oblique-rotator model (blue) using values of $\psi = 0.14^{+0.01}_{-0.08}$, $\gamma = 1.21^{+0.31}_{-0.77}$, and $i = 1.08 ^{+0.40}_{-0.82}$ rad; the best-fit parameters are as determined from the nested-sampling analysis. The phase data for $f_{4,5}$ are shown for sector 12 (orange x). As discussed in the text, the fit requires a value of $\Omega,$ which corresponds to a period of 5.9131 days and is four times slower than the proposed rotation period of the star.  }
\end{figure}

\begin{figure}
    \includegraphics[width=\linewidth]{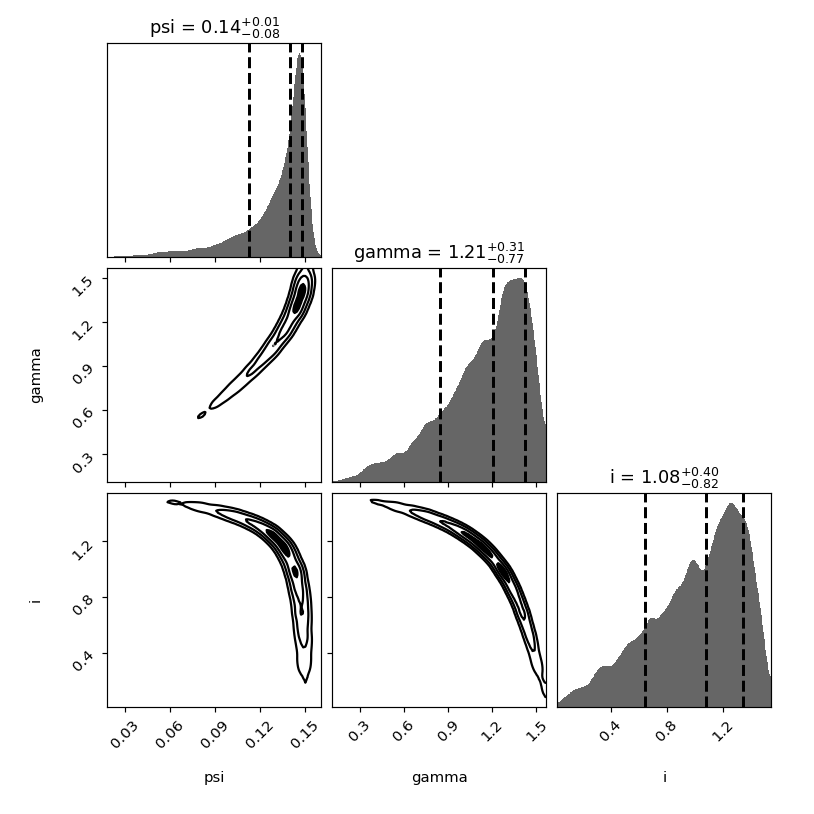}
    \caption{\label{fig:corner}Correlations from our nested sampling of the parameters $\Psi$, $\gamma$, and $i$ for fits to the phase variation of $f_{4,5}$ over sector 12. Best-fit parameters are given above the histograms in radians. }
\end{figure}

The oblique rotator model \citep{Stibbs1950} also predicts that the observed mean longitudinal magnetic field will vary according to the expression  
\begin{equation}
\langle B_z \rangle \propto \cos\beta\cos i + \sin\beta\sin i \cos\Omega t.
\label{bvar}
\end{equation}
\citet{shultz:2021} did not report any strong variation in the magnetic field of HD 156424A. This is not in conflict with the oblique pulsator model, as they report $\beta < 20^{\circ}$, which would make the amplitude of the variation ($\propto \sin \beta$) quite small. The magnetic-field angle for the secondary is also small, at $31^{+20}_{-31}$, so if the variation arises in the secondary, similar results can be expected.

The phase variation in $f_3$ is also smooth, but the period of phase variability is not the same as the period in $f_2$ and $f_{4,5}$;  it also changes from sector to sector, ranging from $P=1.857$ d (sector 66) to $P=2.883$ d (sector 12). None of these periodicities show any relationship with the rotation frequency. 
The last frequency, $f_6$, also shows some short-term phase variability, but relatively large errors in the phase determinations make it difficult to accurately detect any regular variation in the phase.

\subsection{Long-term variability}

We also combined our phase determinations to look for longer term trends in our data. In this case we fitted the phase variability derived using 5.8-day bins, which were chosen to ensure we were averaging over longer time spans than the longest period observed in the short-term phase variability data. We then fitted the resulting phase data for each frequency to a sinusoid, 
\begin{equation}
    \label{eqn:sinfit}
y = a *\sin(2\pi b t + c) + d
,\end{equation}
using a weighted least-squares fit. The strongest variation is observed in $f_2$, which is well fitted by a sinusoidal function with a period of $7400^{+14000}_{-5200}$  d (20.3 years).   
The phase measurements and sinusoidal fit for $f_2$ are shown in Figure~\ref{fig:orbitfit}. The variation in $f_3$ is much longer, with a period of nearly 300,000 years.  The errors and amplitude of the fit are extremely uncertain in this case, and the fit is also consistent with a flat line (a = $22 \pm 1.87\times10^5$).  

\begin{figure}
    \includegraphics[width=\linewidth]{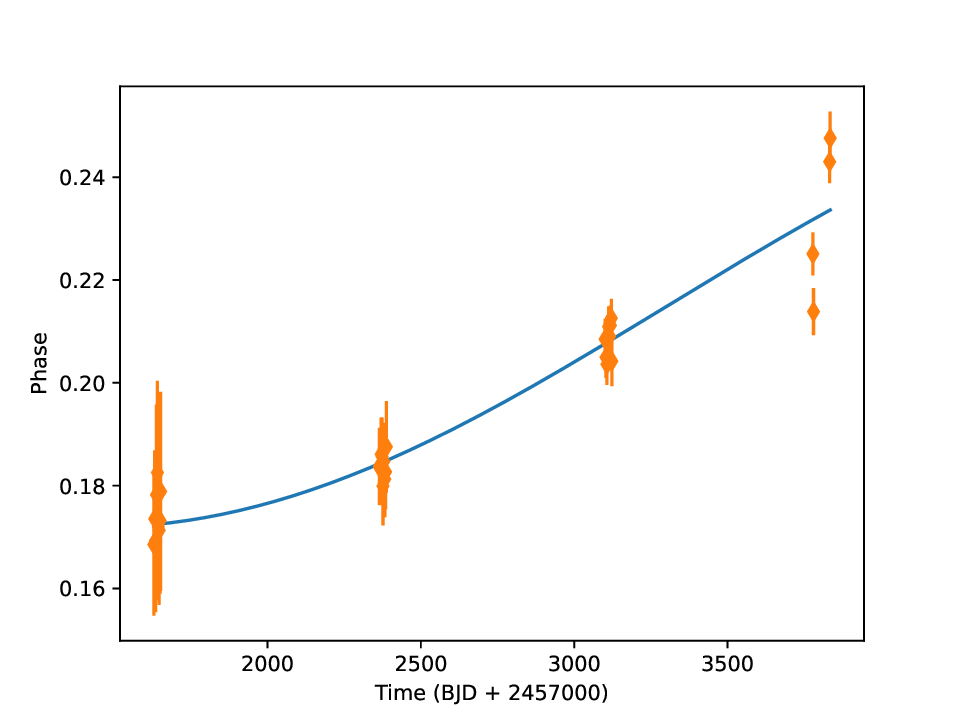}
    \caption{\label{fig:orbitfit}Long-term phase variation in $f_2$ derived using 5.8~d sampling bins fits to a sinusoid (Eq. \ref{eqn:sinfit}). The best fit suggests a long-term variability with a period of 7400 d (20.3 years). }
\end{figure}

The third frequency with phase variability significant enough to check for long-term variability was $f_{4,5}$, which was well fitted by sinusoidal variation and a period of 1490 $\pm$ 10d (4.0 years).  This last period is in closest agreement with the five-year period proposed by \citet{shultz:2021}.

We generated artificial radial-velocity curves for each frequency based on our fits by converting our phase differences to time delays and then taking the derivative \citep[for more details, see][]{murphy2014}. For $f_2$ we found a peak amplitude of $155 \pm 5$ m s$^{-1}$, and $f_3$ gives a maximum radial velocity of about 90 m s$^{-1}$. However, the radial-velocity curve in this case has clearly not reached a maximum value, but it shows a linear change over the period of the TESS observations. For $f_{4,5}$ the amplitude of the radial-velocity curve is much larger, with a peak of 6000 m s$^{-1}$.  In comparison, \citet{shultz:2021} showed radial-velocity variations with amplitude of about $8 \pm 1$ km s$^{-1}$ for the star (their Figure 2). It seems that the parameters we derive for the phase velocity of $f_{4,5}$ are in closest agreement with these previous values. 

 Combined with our best-fit model mass of 7.5 M$_{\odot}$ (see Sect.\ref{sec:results}), a maximum radial velocity of 150 m/s, and a period of 20.3 years (7400 d), iteratively solving the binary mass function gives an upper mass limit of 0.05 $M_{\odot}$ for the proposed third element, assuming $i = 0$ and $ e = 0$.  This would place the companion 15 AU from the primary star. If we assume the higher radial velocity and shorter period associated with $f_{4,5}$ is correct, the companion mass rises to 1.4 M${\odot}$ and is located a distance of 5 AU from the primary. In either case, the companion is of much lower mass than the primary and would not be easily visible in either spectroscopy or photometry.  In fact, if the parameters associated with $f_2$ are correct, the companion may be a brown dwarf rather than a star. Either way, we do not expect the  companion to have a significant effect on the evolution of the primary star.

\section{Asteroseismic models} \label{sec:methods}

To model these stars we used Modules for Experiments in Stellar Astrophysics (MESA) version 15140 \citep{paxton:2011,paxton:2013,paxton:2015,paxton:2018,paxton:2019} with the modified torque and wind routines developed by \cite{keszthelyi:2020} to include the effects of fossil magnetic fields.  The mass loss and stellar winds were also implemented by these routines. As HD 156424 is a rather young star, we selected metallicities at solar and above.  We ran grids at metallicities of Z = 0.014, 0.02, and 0.03 with masses between 7.5 and 10.0 $\textnormal{M}_{\odot}$ based on the estimated mass given by \citet{shultz:2021}. We used a mixing-length parameter of 2.0 and no convective overshoot. Rotation was imposed on the zero-age main-sequence (ZAMS) at rates between 10 and 150 km $\textnormal{s}^{-1}$ and polar magnetic fields with strengths from 3-8 kG, which is consistent with the observed magnetic field strength. The parameter range and step size used in each grid are summarised in Table \ref{tab:parameters}. We used a magnetic braking efficiency of 1.0 and flux conservation.
We also used the uniform torque method, in which angular momentum lost from the stellar surface is removed from the total angular momentum reservoir rather than only the near-surface layers. 

We modelled the primary and secondary separately.  With a projected separation of 450 AU, the two stars are expected to have evolved independently, and there is no need to include the effects of binarity.  The presence of the third low-mass component is also not expected to affect the evolution of the two magnetic stars. Using the lower limit of a 1 M$_{\odot}$ companion in a five-year orbit from \citet{shultz:2021}, 
the semi-major axis is predicted to be around 1.5 AU.  Our results suggest the companion may be even smaller and more distant, with a correspondingly smaller effect on the primary. As such, these stars can safely be modelled as single stars. A subset of our evolution grid is shown in Figure \ref{fig:HR}, along with the observed position of HD 156424A based on values from \cite{shultz:2021}.

\begin{figure}
    \includegraphics[width=\linewidth]{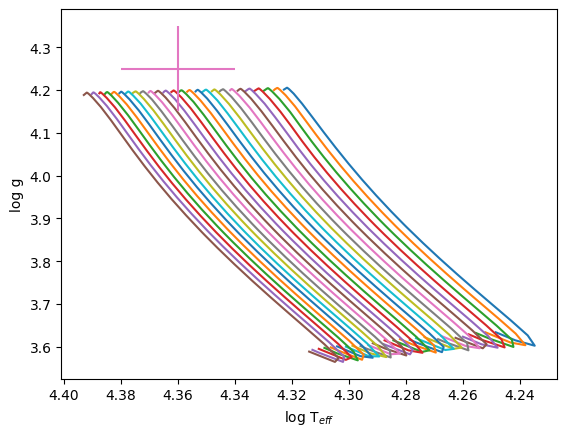}
    \caption{\label{fig:HR}Kiel diagram showing a subset of our model grid. Shown are models with Z = 0.03, v$_{ZAMS}$ = 100 km s$^{-1}$, and $\vec{B} = 12$kG. Tracks shown range from 7.5 M$_{\odot}$ to 10.0 M$_{\odot}$. }
\end{figure}

We saved models every 50 time steps, giving approximately five points along each model's main sequence. We then used GYRE version 7.1 \citep{townsend:2013,townsend:2018,goldstein:2020,sun:2023} to calculate the adiabatic and non-adiabatic pulsation frequencies in the range of 10-25 d$^{-1}$ for each model with $\ell$ = 0, 1, 2, and 3, and $|m| \leq 2 $.   As discussed above, we assumed that $f_1 = 0.718 \rm{d}^{-1}$ is the rotation frequency and did not include it in our asteroseismic modelling.

We then ran a similar grid of models of the secondary star with metallicities Z = 0.02, 0.03, and 0.04; masses from 4.2--6.0 $\textnormal{M}_{\odot}$; and equatorial magnetic field strengths of 3000--8000 G. This corresponds to the range of magnetic fields observed in HD 156424A \citep{shultz:2021}. We calculated non-adiabatic pulsation frequencies in the same way for both sets of models.

\begin{table} [h!]
    \centering
    \begin{tabular}{lcccccc}    
    &\multicolumn{3}{c}{primary}&\multicolumn{3}{c}{secondary} \\
    \hline
    &min &max &step size &min &max &step size \\
    \hline
    $v_{\textnormal{rot}}$\ (km\ $\textnormal{s}^{-1}$) &25 &150 &25\tablefootmark{a} &25 &150 &25 \\
    $M_{\star}$\ ($\textnormal{M}_{\odot}$) &7.0 &10.0 &0.1 &4.2 &6.0 &0.1 \\
    $B_0$\ (kG) &3 &8 &1 &3 &6.5 &0.5 \\
    \hline
    \end{tabular}
    \caption{ Parameters used for the grids of MESA models of the primary and secondary star. \\
    {$^a$}{Grid also includes a model at 10 km s$^{-1}$}}
    \label{tab:parameters}
\end{table}

\section{Asteroseismic fitting} \label{sec:results}

The observed orbital separation of the A and B components of the system is less than 1'' \citep{hartkopf1993,tokovinin2010}, which is well within a single pixel of the TESS observations.  The observed light curve is thus a combination of the light from both stars. Based on our phase-modulation results, we looked at three different cases: i) all frequencies are from pulsation in the primary star; ii) all frequencies are from pulsation in the low-mass secondary star; and iii) the frequency $f_3$ belongs to the secondary star.

First, we assumed that all frequencies ($f_2$, $f_3$, $f_{4,5}$, and $f_6)$ were associated with the primary star.  In this case, we found a total of 28 models with $\chi^2$ values below 0.5, all of which are from the Z = 0.03 model grid. For this case, the Z = 0.014 and 0.02 grids had no models with $\chi^2 < 8$. The model with the lowest $\chi^2$ was from the Z = 0.03 grid: an 8.1 M$_{\odot}$ model with an initial rotation velocity of 25 km s$^{-1}$ on the ZAMS. This was true regardless of the imposed magnetic-field strength. The $\chi^2$ values for models early on the main sequence are identical, and the differences remain small as the models evolve. We imposed the magnetic field at the ZAMS, and so magnetic effects did not have much time to act in the early main sequence, thus, this is perhaps not surprising. In real stars, assuming the magnetic field develops on the pre-main sequence, the effects are likely to be more pronounced.  The low rotation velocity observed in HD 156424A may also indicate that magnetic breaking was significant on the pre-main sequence, leading to minimal effects at the current stage of life. The parameters of the best-fit model are summarised in Table \ref{tab:bestfit_properties}.

Given the model's rotation velocity and radius for our best-fit model, we find a theoretical rotation frequency of 0.133 d$^{-1}$, which is much more slowly rotating than suggested by either $f_1$ or the period of 2.889 implied by the phase variation in Figure \ref{fig:phased}. This model is in good agreement with the observed temperature and luminosity of the star, as shown in Figure \ref{fig:Z03_kiel}. Our best-fit rotation velocity of 25 km s$^{-1}$, combined with the literature measurement of $v\sin i = 4.4$ km s$^{-1}$ \citep{shultz:2021}, implies an inclination of 10.1$^{\circ}$, which would suggest the star is observed nearly pole on. This value is much lower than the best fit derived for the oblique pulsator model in Section \ref{sec:shortterm}.

The comparison to the observed frequencies is shown for our best-ft Z = 0.03 model in Figure \ref{fig:Z03_kiel}.  We see positive excitation in frequencies below about 15 d$^{-1}$, and so we can expect $f_2$, $f_3$, and $f_{4,5}$ to be driven. There are no frequencies with positive growth rates in the range of $f_6$. However, as noted in \citet{shultz:2021}, this frequency is approximately $2f_2$ and is likely not an intrinsic frequency. Removing this frequency from our fitting process does not produce a significant change in the distribution of best-fit models.

\begin{table} [h!]
    \centering
       \caption{Properties of the best-fit models. }
    \begin{tabular} {lcccc}
    &Case 1 &Case 2 &Case 3 &Literature \\
    & & & & values \\
    \hline
    Z & 0.03 & 0.02 & 0.02 \\
    Age (years) &3.10e6 &1.71e6 &8.48e6 \\
    X$_c$ &0.64 &0.70 &0.61 \\
    $M_{\star}$\ ($\textnormal{M}_{\odot}$) &7.5 &4.7 &7.3 &8.8 $\pm$ 0.6 \\
    $v_{\textnormal{ZAMS}}$\ (km\ $\textnormal{s}^{-1}$) &125 &125 &150 \\
    $v_{\textnormal{rot}}$\ (km\ $\textnormal{s}^{-1}$) &107 &120 &97 & 4.4$\pm$1.5\tablefootmark{a} \\
    $R$\ ($\textnormal{R}_{\odot}$) &3.72 &2.61 &3.70 &3.8 $\pm$ 0.2 \\
    $B_0$\ (kG) &2 &any &2.5 &3-16 $\pm$ 50 \\
    log T &4.316 &4.21 &4.312 &4.36 $\pm$ 0.02\\
    log g &4.172 &4.28 &4.164 &4.25 \\
    log L &3.360 &2.63 &3.343 &3.5 \\
    $\chi^2$ & 0.084 & 8.866 & 0.006 & \\
    \hline
    \end{tabular}

     $^a$Published value of $v\sin i$ \citep{shultz:2021}.
 
    {Case 1 assumes that all four $p$-mode frequencies are associated with the primary, Case 2 assumes they are all associated with the secondary, and Case 3 models the primary with only $f_2$ and $f_{4,5}$. Literature values are given for the primary component from \citet{shultz:2021}.}\\

    \label{tab:bestfit_properties} 

\end{table}

\begin{figure}
    \centering
    \includegraphics[width=1\linewidth]{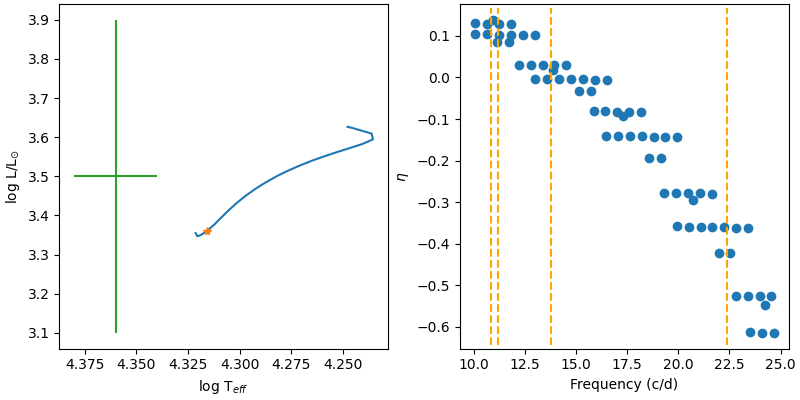}
    \caption{Left: HR diagram of the best-fit Z = 0.03 model of the primary star. The orange star shows the location of the model, and the green error bars indicate the observed location of HD 156424A. Right: Predicted mode excitation for the best-fit model in the 10--25 d$^{-1}$ range. Vertical dashed lines indicate the location of the four primary frequencies from the TESS observations.}
    \label{fig:Z03_kiel}
\end{figure}

Second (Case 2), we compared the observed frequencies with the models of the secondary star (see Fig. \ref{fig:secondary}). Based on the magnetic models of \citet{shultz:2021}, we calculated a grid of models between 4.2 and 6 M$_{\odot}$.   When we compared these models to the observed frequencies, we found that all models of the secondary match poorly, with much higher $\chi^2$ values than the primary models. The secondary star models also have fewer frequencies that are expected to be excited than the primary star models. The best-fit models at each metallicity are shown in Figure \ref{fig:secondary}, and the properties of the best-fit model are summarised in Table \ref{tab:bestfit_properties}.  

Although we modelled the secondary star as a 4-6 M$_{\odot}$ star, it seems likely that it is actually significantly more massive.  \citet{tokovinin2010} cited a magnitude difference of 2.3 mag in the $y$ band, which they noted is probably overestimated. \citet{shultz:2021} used this magnitude difference to estimate that the companion is dimmer by a factor of about eight in terms of luminosity.  However, the visual magnitudes from the Tycho double-star catalogue \citep{Tycho2002} give V-band magnitudes of 9.24 and 9.75 for the primary and the secondary, respectively.  This difference of 0.51 magnitudes implies the primary is a factor of 1.6 brighter than the secondary, giving the secondary $\log L \approx 3.5$.  Similar magnitude differences are seen in Gaia DR3 data \citep{Gaia2016,Gaia2023}. This magnitude difference suggests the secondary is closer to 7.5--8 M$_{\odot}$ rather than the 5 M$_{\odot}$ suggested by \citet{shultz:2021}.  This mass range is more consistent with our primary model grid, and so we did not calculate additional models for the secondary in this mass range. If this mass of the secondary is correct, it is equivalent to the models in Case 1. However, based on ESPaDOnS spectra, \citet{shultz:2021} found that the pattern of ionisation lines is consistent with a much cooler companion, with the effective temperature of the primary at 23,000 K and the secondary at 16,000 K; this is more consistent with the lower mass range.

\begin{figure}
    \centering
    \includegraphics[width=1\linewidth]{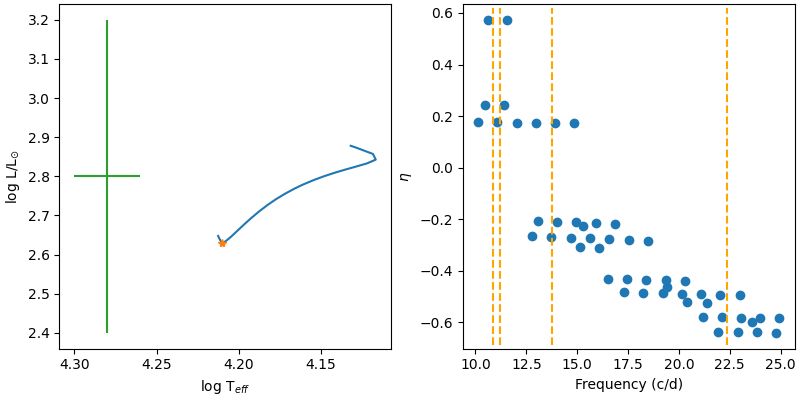}
    \caption{As for Figure \ref{fig:Z03_kiel} but for best-fit model of the secondary star with Z = 0.02. The best-fit model is 4.7 M$_{\odot}$.}
    \label{fig:secondary}
\end{figure}

Finally (Case 3), we tried dividing up the frequencies between the two stars based on our phase data in Section \ref{sec:shortterm}.  Since we see beating between $f_2$ and $f_{4,5}$, we assumed that these two frequencies must be associated with the same star. Since $f_6$ appears to be $2f_2$, we included it as well, although excluding it would not significantly change our results. This leaves only $f_3 = 13.7 \rm{d}^{-1}$ as being potentially associated with the secondary. With only one frequency as a constraint, we did not attempt to fit models of the secondary in this case. 

When we fitted the models in the primary grid to frequencies $f_2$, $f_{4,5}$, and $f_6$, we found that our best-fit model is slightly less massive and more slowly rotating than in Case 1, as shown in Table \ref{tab:bestfit_properties}.   This model is more rapidly rotating, and the rotation frequency is expected to be 0.5180 d$^{-1}$. As with the best-fit model in Case 1, this rotation frequency does not correspond to any of the frequencies observed in our data. The inclination of the model in this case would be 1.3$^{\circ}$, i.e. in much worse agreement with the value derived from phase fits in Section \ref{sec:phases}.

\begin{figure}
    \centering
    \includegraphics[width=1\linewidth]{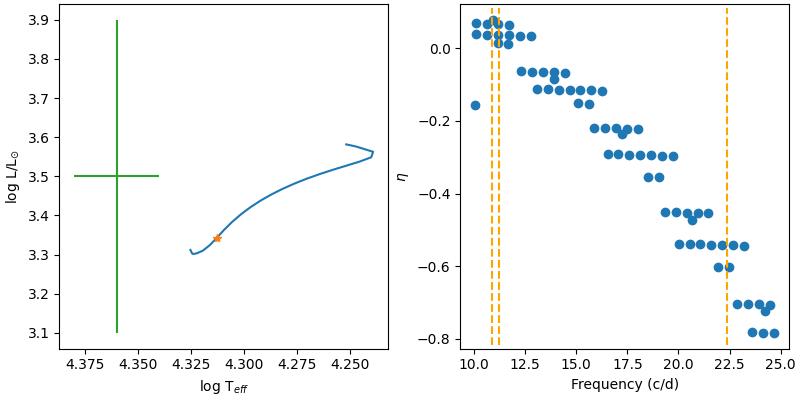}
    \caption{Same as Fig. \ref{fig:Z03_kiel} but for the primary star in Case 3.}
    \label{fig:case3}
\end{figure}

Overall, our models suggest that the HD156424 system is quite young, and we find that modes in the observed range of p-mode frequencies are not predicted to be excited in the later part of the main sequence for stars in this mass range. Higher metallicity models (Z=0.03) give better fits than the lower metallicity models (Z = 0.014 and 0.02), although our best-fit individual model in Cases 2 and 3 comes from the Z = 0.02 grid. None of our best-fit models rotate at the proposed rotation frequency, and the derived inclinations are in poor agreement with the value derived from the phase variation in Section \ref{sec:phases}.

\section{Conclusions}\label{sec:conclusions}

We analysed five sectors of TESS observations for HD 156424, both individually and as a combined dataset. We found seven 
significant frequencies across all sectors and five frequencies that are significant in individual sectors. Of our detected frequencies, the pair $f_4 $ and $ f_5$ are extremely close, and they likely indicate frequency variability over the course of the TESS observations.  For $f_{4,5}$, the amplitude is strongly variable between and within sectors, as shown in Figure \ref{fig:river}. We conclude that $f_1 = 0.7171$ d$^{-1}$ likely corresponds to the rotation frequency of the primary, HD~156424A. 

Following the method of \citep{murphy2014}, we divided our light curve into 0.64 d bins and calculated the phase of each pulsation frequency in each bin. We found significant short term variability in $f_2$, $f_3$, and $f_{4,5}$. Both $f_2$ and $f_{4,5}$ have phase modulation with a period of 2.889 d, which is close to, but not exactly, $f_1/2$. We suggest that this may be the result of differential rotation in the envelope of the star, resulting in a slightly longer rotation period in the driving region of the $p-$modes. Since the driving region in these modes is quite close to the surface, the differential rotation is probably at the surface, with the poles rotating more rapidly than the equator. 

We found smooth variation for $f_2$, which suggests that the variation is caused either by beating or by differential rotation between the driving regions of $f_2$ and $f_{4,5}$. The observed beat period of 2.889 d corresponds to the frequency 0.346 d$^{-1}$, which is approximately the separation between $f_2$ and $f_{4,5}$.  If the variation arises from differential rotation, the angular momentum profile required would be unusual, with the the sub-surface region rotating more slowly than the surface.  For this reason, beating between the two modes seems to be a simpler and more likely explanation. The variation in $f_{4,5}$, although it has the same period as the variation in $f_2$, shows very different behaviour. There is a gradual increase over each cycle, followed by a sharp drop.  This phase behaviour is consistent with the expected behaviour of an oblique magnetic pulsator \citep{bigot2002,Bigot2011}. We were able to fit our data from sector 12 to this model and found reasonable fits with $\psi = 0.14^{+0.01}_{-0.08}$, $\gamma = 1.21^{+0.31}_{-0.77}$, and $i = 1.08 ^{+0.40}_{-0.82}$ radians. Fits to the data from sector 66 show agreement within the errors.  

We then divided our light curve into 5.8 day bins to look for long-term variability. We found strong long-term variations in $f_2$ and $f_{4,5}$. The period in the phase variation of $f_2$ was $7400^{+14000}_{-5200}$ d.
The long-term variability of $f_2$ is strongly modified by the data derived from sector 93. The four\ points of binned data for this sector significantly increase the error bar for the derived period of the phase variability. Further TESS observations of HD 156424 may improve this estimate.

Our derived variability period for $f_{4,5}$ was in better agreement with previous data, with a period of $1490 \pm 10$ d.  The phase variation in $f_3$ was poorly constrained, and our fit could also be consistent with a flat line. In all cases the errors on the phase calculations were large, and further monitoring of this system will be required to confirm and refine the period of the proposed third element. 

We used the long-term phase variations to derive radial-velocity curves following the method outlined in \citet{murphy2014}. The resulting radial velocity for $f_2$ was around 155 m s$^{-1}$, while $f_{4,5}$ gave a much higher radial velocity of 8000 m s$^{-1}$, which is in better agreement with the literature values \citep{shultz2018}. The radial velocities for $f_3$ are extremely low, with a maximum in our dataset of 90 m s$^{-1}$. However, the variation is linear, and we are likely underestimating the maximum radial velocity in the system. Using our new values for the periods and radial velocities results in a much smaller object as a third companion, with an upper limit of 0.05 M$_{\odot}$ derived from the $f_2$ fit. As this is an upper limit, the companion could well be a brown dwarf or a hot Jupiter.  

Our asteroseismic models show the system is quite young, almost at the ZAMS, which is consistent with its membership in the Sco OB4 association. This is true across all metallicities we considered.  Our best-fit model has a mass of 7.3 $M_{\odot}$ with metallicity Z = 0.02, a surface temperature of $\log T = 4.312$, a surface gravity of $\log g = 4.164$, and luminosity of $\log L$ = 3.343 (see Table~\ref{tab:bestfit_properties}, Case 3). The rotation velocity of this model is 97 km s$^{-1}$, which, combined with the radius of 3.822 R$_{\odot,}$ gives a rotation frequency of 0.5180 d$^{-1}$.

The A and B components of HD 156424 are within 1'' on the sky, so we also considered the possibility that the pulsations originate in the lower mass secondary.  However, we found no good fits to the observed frequency between 4 and 6 M$_{\odot}$.  The mass of 5 M$_{\odot}$ is based on a luminosity difference of 2.3 mag in the $y$ band \citep{tokovinin2010}.  However, the Gaia DR3 magnitudes and the B and V magnitudes from the Tycho Double Star Catalogue \citep{Tycho2002} are much closer and suggest the secondary has a mass of 8 M$_{\odot}$, which is more similar to that of the primary star.  

Regardless of the mass of the secondary, our asteroseismic fitting suggests a young age for the system. This is an interesting result, as much of the previous work suggests that the fossil magnetic fields seen in the upper main sequence are the result of merger processes \citep{ferrario2009,keszthelyi2021}. While \citet{keszthelyi2021} and others suggest that $\tau$ Sco has been rejuvenated by a recent merger, \citet{ferrario2009} argued that magnetic fields can be produced by a merger of two protostars, driving strong differential rotation and hence a strong dynamo. Given the characteristics of HD~156424 derived here, a merger scenario seems unlikely, suggesting magnetic fields in massive stars can have other origins.

\begin{acknowledgements}
This work was supported by an NSERC Discovery grant to CCL.  V.K. acknowledges support from Mitacs, ACFAS, and is thankful to the Facult\'{e} des \'{E}tudes Sup\'{e}rieures et de la Recherch and to the Facult\'{e} des Sciences de l'Universit\'{e} de Moncton for financial support of this research. Computational resources for this project were provided by the Digital Research Alliance of Canada. TESS data was obtained from the MAST archive at
https://dx.doi.org/10.17909/T9RP4V
 This work has made use of data from the European Space Agency (ESA) mission
{\it Gaia} (\url{https://www.cosmos.esa.int/gaia}), processed by the {\it Gaia}
Data Processing and Analysis Consortium (DPAC,
\url{https://www.cosmos.esa.int/web/gaia/dpac/consortium}). Funding for the DPAC
has been provided by national institutions, in particular the institutions
participating in the {\it Gaia} Multilateral Agreement. The authors would also like to acknowledge the referee, who provided helpful comments that improved the paper.
\end{acknowledgements}

  \bibliography{aa56715-25corr}{}
\bibliographystyle{aa}

\end{document}